\documentstyle[psfig,multicol]{article}
\newcommand{\al}{{\em et al }}
\newcommand{\eg}{{\em e.g. }}

\newcommand{\mylist}{\begin{list}{$\diamondsuit$}{\setlength{\leftmargin}{0cm}}}

\begin{document}

\title{Coupled Two-Way Clustering 
Analysis of Gene Microarray Data}
\author{G. Getz, E. Levine and E. Domany \\
{\small Department of Physics of Complex Systems,} \\
{\small Weizmann Inst. of Science, Rehovot 76100, Israel}}
\date{\today}
\maketitle
\begin{abstract}
We present a novel {\em coupled} two-way clustering approach 
to gene microarray data analysis. 
The main idea is to identify subsets of the genes and samples,
such that when one of these is used to cluster the other, 
stable and significant partitions emerge. 
The search for such subsets is a computationally complex task: 
we present an algorithm, based on iterative clustering, which 
performs such a search. 
This analysis is especially suitable
for gene microarray data, where the contributions of 
a variety of biological mechanisms 
to the gene expression levels are entangled in a large body 
of experimental data. The method was applied
to two gene microarray data sets, on colon cancer and leukemia.
By identifying relevant subsets of the data and 
focusing on them we were able to discover
partitions and correlations that were masked and hidden when 
the full dataset
was used in the analysis. Some of these partitions have clear biological
interpretation; others can serve to identify possible directions for 
future research.
\end{abstract}

\begin{multicols}{2}
\section*{Introduction}

In a typical DNA microarray experiment expression levels of thousands of genes 
are recorded
over a few tens of different samples\footnote{
By ``sample" we refer to any kind of living matter that is being tested, e.g.
different tissues\cite{Alon99}
cell populations collected at different times\cite{Eisen98} etc. }
\cite{Alon99,Golub99,Perou99}. 
Hence this new technology gave rise to a new computational
challenge:  to make sense of such massive expression data 
\cite{Lander99,Zhang99,Marcotte99}.
The sizes of the datasets and their complexity call for multi-variant
clustering techniques \cite{Hartigan75,Kohonen97}, 
which are essential for extracting correlated patterns and the natural 
classes present in  a set of $N$ data points, or {\it objects}, represented 
as points in the multidimensional space defined by $D$  measured {\it features}.

Gene microarray data are fairly special in that it makes good sense to perform
clustering analysis in two ways~\cite{Alon99,Eisen98}. The
first views the $n_s$ {\it samples} as the $N=n_s$ objects to be clustered,
with the $n_g$ genes' levels of expression in a particular sample playing
the role of the features, representing that sample as a point in 
a $D=n_g$ dimensional space.
The different phases of a cellular process emerge from grouping together samples 
with similar or related expression profiles. 
The other, not less natural way, looks for clusters of  genes 
that act correlatively on the different samples. This view considers the $N=n_g$
{\it genes} as the objects to be clustered, each represented by
its expression profile, as measured over all the samples, as a point in a
$D=n_s$ dimensional space. 

Whereas in previous work~\cite{Alon99, Eisen98, Alizadeh00} the samples
and  genes were clustered completely independently, we introduce and perform here 
a {\em coupled} two-way clustering (CTWC) analysis. 

Our philosophy is to narrow down both the features that
we use and  the data points that are clustered. We believe that 
only a small subset of the genes participate in any cellular process of 
interest, which takes place only in a subset of the samples;  by focusing 
on small subsets,  we lower the noise induced by the other samples and genes.
We look for pairs of a relatively small subset $\cal F$ of features 
(either genes or samples) and of objects  $\cal O$, (samples or genes), such that when
the set $\cal O$ is clustered using the features $\cal F$, stable and significant
partitions are obtained. Finding such pairs of subsets is
a rather complex mathematical
problem; the CTWC method produces such pairs in an iterative clustering process.

CTWC can be performed with any clustering algorithm.
We tested it in conjunction with
several clustering methods, but present here only results that were obtained 
using the {\em super-paramagnetic}
clustering algorithm (SPC) \cite{Blatt96,Domany99,Getz00}, which
is especially suitable for gene microarray data analysis due
to its robustness against noise and its ``natural'' ability to identify
stable clusters. 

The CTWC  clustering scheme was applied to two gene microarray
data sets, one from a colon cancer experiment \cite{Alon99} and the
other from a leukemia experiment \cite{Golub99}.
From both datasets we were able to ``mine" new  partitions 
and correlations that have {\it not} 
been obtained in an unsupervised fashion by previously used methods.
Some of these new partitions have clear, well understood biological
interpretation. 
We do {\it not} report here 
discoveries of biologically relevant, previously unknown results.
The main point of our message is twofold: (a) we 
{\it were} able to identify biologically relevant 
partitions in an unsupervised way and
(b) other, not less natural new partitions were also found,
which {\it may} contain new, important information and
for which one should seek biological interpretation.


\section*{Coupled Two Way Clustering}

\subsection*{Motivation and Algorithm}
\label{sub:algorithm}
The results of every gene microarray experiment can be summarized as
a set of numbers, which we 
organize in an {\em expression level matrix} ${\cal A}$. 
A row of this matrix corresponds to a single gene,  
while each column represents a particular sample. 
Our normalization is described in detail later. 

In a typical experiment simultaneous expression levels of thousands of
genes are measured. Gene expression is influenced by
the cell type, cell phase, external signals and more \cite{Alberts94}.
The expression level matrix is therefore the result of all
these processes mixed together. 
Our goal is to separate and identify these processes
and to extract as much information as possible 
about them. The main point is that each biological process 
on which we wish to focus may involve a relatively
small subset of the genes that are present on a microarray; the large majority
of the genes constitute a noisy background which may mask the effect of the
small subset. The same may happen with respect to samples. 

The CTWC procedure which we now describe is designed to identify
subsets of genes and samples, such that a single process is the main
contributor to the expression of the gene subset over the sample subset.
We start with clustering the samples and the genes 
of the full data set and identify all stable clusters of either samples or genes.
We scan these clusters one by one. The 
expression levels of the genes of each cluster are used 
as the feature set $\cal F$ to represent object sets.  
The different object sets $\cal O$ contain either all
the samples or any sample cluster.
Similarly, we scan all stable clusters of samples
and use them as the feature set $\cal F$ to identify stable clusters of genes.
We keep track of all the stable clusters that are
generated, of both genes, denoted as $v^g$, and samples $v^s$. 
The gene clusters are accumulated in a list $V^g$ and the sample clusters in
$V^s$. Furthermore, we keep all the chain of clustering analyses that has been 
performed (which subset was used as objects, which subset was used as features,
and which were the stable clusters that have been identified).

When new clusters are found, we use them in the next iteration.
At each iteration step we cluster 
a subset of the objects (either samples or genes) using a subset of
the features (genes or samples). The procedure stops when no new 
relevant information is generated. 
The outcome of the CTWC algorithm are the final sets $V^g$ and $V^s$ and the
pointers that identify how all 
stable clusters of genes and samples were generated.

A precise, step by step definition of the algorithm is given 
in Fig. \ref{fig:algorithm}.

\begin{figure*}[tb]
{\small  \centerline{
\begin{tabular}{|cp{7.0cm}|}
\hline
 {\bf Step 1.} &  {\bf Initialization} 
   \\ {\bf 1a.} &  Let $v^g_0$ be the cluster of all genes, and $v^s_0$ be
      the cluster of all samples. 
   \\ {\bf 1b.} &  Initialize  sets 
      of gene clusters, $V^g$, and sample clusters, $V^s$, 
	  such that  $V^g=\{v^g_0\}$ and 
      $V^s=\{v^s_0\}$. 
   \\ {\bf 1c.} &  Add each known class of genes as a member of $V^g$, 
      and each known class of samples as a member of $V^s$.
   \\ {\bf 1d.} &  Define a new set $W=\emptyset$. This set is needed to keep track of
      clustering analyses that have already been performed.
\\&
   \\ {\bf Step 2.} &  For each pair $(v^g,v^s) \in \left( V^g \times V^s \right) \setminus W$:
   \\ {\bf 2a.} & Apply the clustering algorithm on the genes of $v^g$ using the samples
      of $v^s$ as its features and vice versa.
   \\ {\bf 2b.} &  Add all the robust gene clusters generated by {\bf Step\,2a}
      to $V^g$, and all the robust sample clusters to $V^s$. 
   \\ {\bf 2c.} & Add $(v^g,v^s)$ to $W$. 
\\&\\ {\bf Step 3.} &  For each new robust cluster $u$ in either $V^g$ or $V^s$ 
      define and store a pair of labels $P_u = (u_o,u_f)$. Of these, $u_o$ 
	  is the cluster of objects which were clustered to find $u$, 
	  and $u_f$ is the cluster of features used  in that clustering.
\\&\\ {\bf Step 4.} &  Repeat {\bf Step\,2} until no new clusters are added 
      to either $V^g$ or $V^s$.
\\ \hline
\end{tabular}
}} 
\caption{\small{CTWC algorithm. The input of the algorithm is the full expression
matrix. The output is a set $V^g$ of stable gene clusters and a set $V^s$
of stable sample clusters. For each stable cluster $u$, found in a clustering 
operation, the clusters which
provided the objects and those that served as the features for this operation 
are stored as a label $P_u$.}}
\label{fig:algorithm}
\end{figure*}


\subsection*{Analyzing the clusters obtained by CTWC}

\label{sec:analyzing}

The output of CTWC has two important components. First, it provides a 
broad list of gene and sample clusters. Second, for each cluster (of samples,
say) we know  which subset (of samples) was clustered to find it, and
which were the
features (genes) used to represent it.  
We also know for every cluster $\cal C$, which other clusters can be 
identified by using $\cal C$ as the feature set.
We present here a brief selection of the possible ways one can
utilize this kind of information. Implementations of the particular 
uses listed here are described in the Applications section.

{\bf Identifying genes that partition the samples 
according to a known classification. } 
This particular application is {\it supervised}. 
Denote by $C$  a known classification
of the samples, say into two classes, $c_1$ and $c_2$. 
CTWC provides an easy way to rank the clusters of genes in
$V^g$ by their ability to separate the samples according to $C$.
It should be noted that CTWC not only provides a list of candidate gene clusters
one should check, but also a unique method of testing them.

First we 
evaluate for each cluster of samples $v^s$
in $V^s$ two scores,
{\it purity} and {\it efficiency}, which 
reflect the extent to which assignment of the samples to $v^s$ 
corresponds to the classification $C$.
These figures of merit are
defined (for $c_1$, say) as 
\[
\mbox{purity}(v^s \vert c_1)=\frac{\vert v^s \cap c_1 \vert }{\vert v^s \vert} \,\,\,; \,\,\,
\mbox{efficiency}(v^s \vert c_1)=\frac{\vert v^s \cap c_1 \vert}{\vert c_1 \vert} .
\]
Once a cluster $v^s$ with high purity and efficiency has been found, we can 
use the saved pointers to read off
the cluster (or clusters) of genes that were used as the feature set 
to yield $v^s$ in our clustering
procedure. 
Clustering, as opposed to
classification, discovers only those partitions of the data which
are, in some sense, ``natural''. Hence by this method we identify the most
natural group of genes that can be used to induce a desired classification.

Needless to say, one can also test a gene cluster $v^g$ that was 
provided by CTWC using
more standard statistics, such as the {\em t-test} \cite{Wadsworth60}
or the Jensen-Shannon distance \cite{Cover91}. Both compare the
expression levels of the genes
of $v^g$ on the two groups of samples, $c_1, c_2$, partitioned according to $C$.
Alternatively, 
one can also use the genes of $v^g$ to train a classifier to
separate the samples according to $C$ \cite{Golub99}, and use 
the success of the classifier to measure whether
the expression levels of 
the genes in $v^g$ do or do not correspond to the classification.

{\bf Discovering new partitions.} 
Every cluster $v^s$ of $V^s$ is a subset of all the samples,
the members of which have been  linked to each other 
and separated from the other
samples on the basis of the expression levels of 
some co-expressed subset of genes. It is reasonable
therefore to argue that the cluster $v^s$ has been formed for some
biological or experimental reason. 

As a first step to understand the reason
for the formation of a robust cluster $v^s$, one should
try to relate it to some previously known classification
(for example, in terms of purity and efficiency). 
Clusters which cannot be associated with any known classification,
have to be inspected more carefully.
Useful hints for the meaning of
such a cluster of samples may come from the identity of the 
cluster of genes
which was used to find it. Clearly, the CTWC clusters can be used in the same way
to interpret clusters of genes which were not previously known to belong to the
same process.

{\bf  CTWC is a sensitive tool to identify sub-partitions.}
Some of the sample clusters in $V^s$ may have have emerged from clustering a
subset of the samples, say $v^s_0$. 
These clusters reflect a sub-partition of the samples which 
belong to $v^s_0$. When trying to cluster the full sample set,
this  sub-partition may be missed,  since other samples, unrelated to
$v^s_0$, are masking it.

{\bf  CTWC reveals conditional correlations among genes.}
The CTWC method collects stable gene clusters in $V^g$. In many cases
the same groups of genes may be added to $V^g$ more than once. 
This is caused by the fact that some genes
are co-regulated in
all cells, and therefore are clustered together, no matter which subset of the 
samples is used as the feature set. 
For example, ribosomal proteins are
expected to be clustered together for any set of samples which is not
unreasonably small.

Some gene clusters, however, are different; they are
co-regulated only in a specific subset of samples. We call
this situation conditional correlation. The identity of the sample
cluster which reveals the conditionally correlated gene cluster is
clearly important to understand the biological process
which  makes these genes correlated.

\section*{Clustering method and similarity measures}
\label{sec:spc}

Any reasonable choice of clustering method and definition of stable clusters
can be used within the framework of CTWC. 
We describe here the benefits of the particular clustering algorithm and similarity
measure we used, which we found to 
be particularly suitable to handle the special properties of
gene microarray data.

\subsection*{SPC provides clear identification of stable clusters in a robust manner.}
{\em Super-paramagnetic} clustering (SPC) is  
a hierarchical clustering method recently introduced by 
Blatt \al \cite{Blatt96}. The intuition that led to it 
is based on an analogy to the physics of inhomogeneous ferromagnets. 
Full details of the algorithm and the
underlying philosophy are given elsewhere~\cite{Domany99,Blatt97}.

As for many hierarchical clustering algorithms, the input
for SPC is a distance or similarity matrix $d_{ij}$ between the objects $\cal O$,
calculated according to the feature set $\cal F$.
A tunable parameter $T$ ('temperature') controls the resolution
of the performed clustering. One starts at $T=0$, with 
a single cluster that contains all the objects. As $T$ increases, phase transitions take place, 
and this cluster breaks into several sub-clusters which reflect the
structure of the data.  Clusters keep breaking up as $T$ is further increased,
until at high enough values of $T$ each object forms its own cluster.

Blatt \al showed that the SPC algorithm is robust against variation of its
parameters, initialization and {\it against noise in the data}. 
The following advantages of SPC makes it especially suitable for gene
microarray data analysis:
({\it i}) No prior knowledge of the structure of the data is assumed;
({\it ii}) SPC provides information about the different self organizing regimes
of the data; ({\it iii}) The number of ``macroscopic'' clusters is an {\em output}
of the algorithm; and ({\it iv}) Hierarchical organization of the data is reflected
in the manner clusters merge or split when the control parameter (the
'temperature' $T$) is varied. 

Moreover, the control parameter can be used to 
provide a natural measure for the stability of any particular cluster by the 
range of temperatures $\Delta T$ at which the cluster remains unchanged. 
A stable cluster is expected to 'survive' throughout a large $\Delta T$, 
one which
constitutes a significant fraction of the range it takes the data to break
into single point clusters. Inspection of the gene dendrograms of 
Fig. \ref{fig:colongenes} reveals stable
clusters and stable branches. 


\subsection*{Normalization of the gene expression array}
\label{sec:normalization}
The Pearson correlation is commonly used as the similarity measure 
between genes or samples \cite{Schena96,Eisen98,Alon99}. 
This measure conforms with the intuitive biological
notion of what it means for two genes to be co-expressed; 
this statistic captures similarity of the ``shapes" of two
expression profiles, and ignores differences between the 
magnitudes of the two series of measurements \cite{Eisen98}. 
The correlation coefficient is high between two genes that are affected 
by the same process, even if each has a different 
gain due to the process, over different background
expression levels (caused by other processes).
One problem of using the correlation coefficient is that its 
reliability depends on the absolute expression level of the 
compared genes; a positive correlation between two 
highly expressed genes is much more significant than the same value 
between two poorly expressed genes. This information is ignored in
the clustering process.

However, we find that correlations do not always 
capture similarity between samples.
For example, consider two samples taken at different stages of some process,
with the expression levels of a family of genes much below average in one sample
and much higher in the other. Even if the expression levels of the two samples 
over these genes are correlated, one would like to assign them into different
clusters.
Furthermore, the distance between the two samples should
be affected by the statistical significance of their expression differences.

We therefore  used the following normalization scheme. Denote by  ${\cal D}$
the matrix of the raw data. 
${\cal D}$ is a $n_g \times n_s$ matrix, where $n_g$ is the number
of genes and $n_s$ the number of samples.

We normalize our expression level matrix in two steps. 
First, divide each {\it column}
by its mean: ${\cal D'}_{ij}={\cal D}_{ij}/{\bar {\cal D}}_{j}$~; 
${\bar {\cal D}}_{j} = \frac{1}{n_g}\sum_{i=1}^{n_g}{{\cal D}_{ij}}$. We then
normalize each row, such that its mean vanishes and its norm is one: 
\[ 
{\cal A}_{ij} = \frac{{\cal D'}_{ij}-{\bar {\cal D'}}_{i}}{\|{\cal D'}_{i}\|},
\]
where ${\bar {\cal D'}}_{i} = \frac{1}{n_s}\sum_{j=1}^{n_s}{{\cal D'}_{ij}}$ and 
$\|{\cal D'}_{i}\|^2 = \sum_{j=1}^{n_s}{\left({\cal D'}_{ij}-{\bar {\cal D'}}_{i} \right)^2 }$.

For genes and samples we use the Euclidean distance as the dissimilarity measure.
For two genes (rows of ${\cal A}$) the
Euclidean distance is closely related to 
the Pearson correlation between them.
\section*{Applications}
\label{sec:app}


In order to show the strength of the CTWC algorithm, we apply it to
two gene microarray experiment data sets. Here we report only the
results which were obtained by CTWC, and {\em could not} be found using
a straightforward clustering analysis. 
We highlight a small subset of the partitions that our method was able to
extract from the data; these are the results for which we were able
to find satisfactory biological explanation. We do {\it not} report here 
new discoveries of biologically relevant, previously unknown results. Rather,
we claim to have discovered a method that is capable to {\it mine} such information
out of the available data. New, relevant information {\it may} be contained
in the new partitions which were found, 
to which we were not yet able to assign
biological meaning. These new, uninterpreted results are reviewed briefly below;
full lists of the clusters associated with these results, as well as their 
constituent samples or genes can be found at 
http://www.weizmann.ac.il/physics/complex/compphys.


\subsection*{Analysis of Leukemia samples}
We analyzed data obtained by Golub \al \cite{Golub99}  
from 72 samples collected from acute leukemia patients at the time
of diagnosis. 
47 cases were diagnosed as ALL (acute lymphoblastic leukemia)
and the other 25 as AML (acute myeloid leukemia).
RNA prepared from the bone marrow mononuclear 
cells was hibridized to high-density oligonucleotide micorarrays, 
produced by Affymetrix, containing 6817 human genes.

\begin{figure*}[tb]
  \centerline{
  \psfig{figure=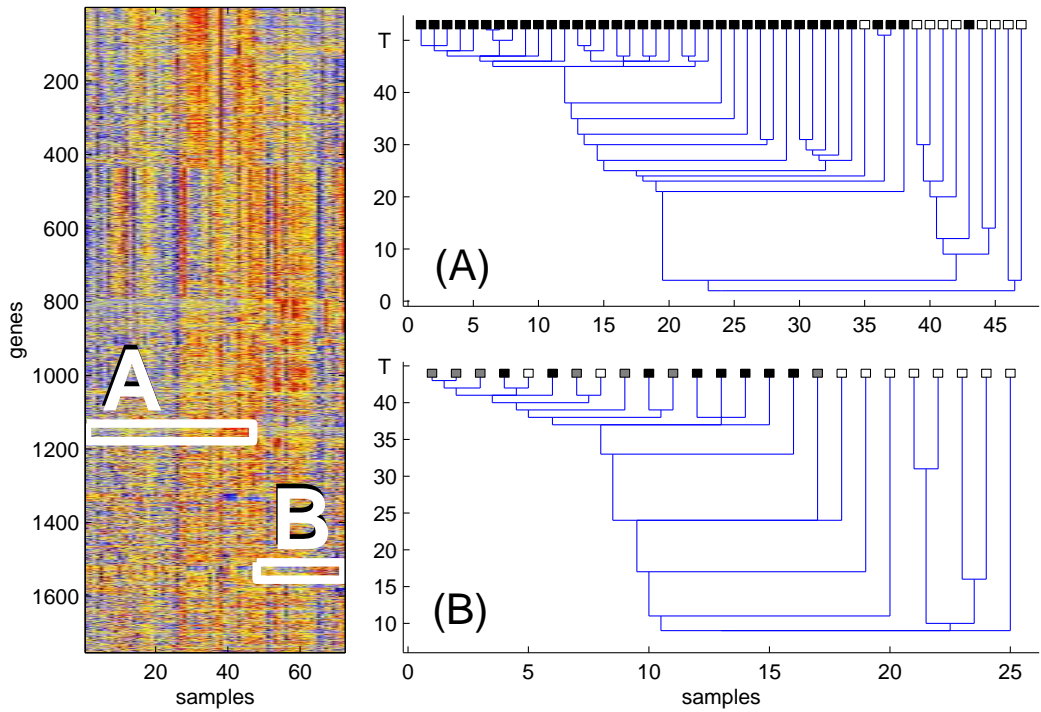,height=9.6cm}
  }
\caption{\small{The expression level matrix of the leukemia
experiment is shown on the left.  Rows correspond to different genes,
ordered by clustering them using all the samples.
The two boxes contain expression data from (A) ALL patients, 
measured on one gene cluster and (B) AML patients, on another gene cluster.
Clustering the ALL samples, using the data in box (A),
yields good separation between T-ALL (black) and B-ALL (white). 
Clustering of AML samples, using the data in box (B) yields a stable cluster, 
which contains all
patients who were  treated, with results known to be either success (black) or failure (gray).
The vertical axis is the ``temperature'' parameter $T$ and on the horizontal axis the samples
are ordered according to the dendrogram.
}}
\label{fig:leukemia}
\end{figure*}

After rescaling the data in the manner described by Golub \al, 
we selected only those genes whose minimal expression over all
samples is greater than 20. As a result of this thresholding operation 
1753 genes were left. The resulting array  
was then normalized as described previously,
to give the $1753 \times 72$ expression level matrix ${\cal A}$
(see Fig. \ref{fig:leukemia}).

We found that two iterations of the
CTWC algorithm
sufficed to converge to  
49 stable gene clusters (LG1-49) and 35 stable sample clusters
(LS1-35). We highlight here four of our findings, 
which demonstrate the power of
the method to solve problems listed above.

{\bf Identifying genes that partition the samples 
according to a known classification. } 
First we use the known ALL/AML classification of the samples to 
determine which gene clusters can distinguish between the two classes. 
We found only a single gene cluster (LG1) which enables
stable separation into AML/ALL clusters\footnote{A cluster is identified with a 
certain class if both its purity and efficiency exceeds 3/4.}.
This well demonstrates the strength of CTWC, since it turned out that SPC
was not able to clearly identify the AML/ALL separation using the full set of genes.

{\bf Discovering new partitions.} 
Next, we search the stable sample clusters for 
unknown partitions of the samples. We focus our attention on sample
clusters which were repeatedly found to be stable.
One such cluster, denoted LS1, may be of interest; it includes 37 samples  
and was found to be stable when either a cluster of 27 genes (LG2) or 
another unrelated cluster of 36 genes (LG3) was used to provide the features.
LG3 includes many genes that participate in
the glycolysis pathway. Due to lack of additional information
about the patients we cannot determine the biological origin 
of the formation of this sample cluster.

{\bf Identifying sub-partitions}
Using a 28 gene cluster (LG4) as features, we tried to cluster only the
samples that were identified as 
AML patients (leaving out ALL samples).  
A stable cluster, LS2, of 16 
samples  was found (see Fig. \ref{fig:leukemia}(B)); 
it contains most of the samples (14/15) 
that were taken from patients that underwent treatment and whose
treatment results were known  (either success or failure).
For none of the other AML patients was any information about treatment
available in the data. 
Some of the 16 genes of this cluster, LG4, 
are ribosomal proteins and some others are related
to cell growth. Apparently these genes 
can partition the AML patients according to whether they did or
did not undergo treatment.

This result demonstrates a possible diagnostic use of the CTWC
approach; one can identify different responses to treatment,
and the groups of genes to be used as the appropriate probe. 

We repeated the same procedure, but discarding AML and keeping 
only the ALL samples. We discovered that when any one of
5 different gene clusters (LG4-8)
are used to provide the features, the ALL samples break into two stable  
clusters; LS5, which consists mostly of T-Cell ALL patients and LS4, that
contains mostly B-Cell ALL patients (see Fig. \ref{fig:leukemia}(A)).
When all the genes were used to cluster 
all samples, no such clear separation into T-ALL vs B-ALL was observed. 
One of the gene clusters used, LG5, with T/B separating ability, contains
29 genes, many of which are T-cell related. 
Another gene cluster, LG6, which also gave rise to T/B differentiation 
contains many HLA histocompatability genes.

These results demonstrate how CTWC can be used to characterize different types of
cancer. Imagine that the nature of the sub-classification of ALL had not been
known. On the basis of our results we could predict that there are two distinct
sub-classes of ALL; moreover, by the 
fact that many genes which induce separation into these
sub-classes are either T-Cell related or HLA genes, one could 
suspect that these sub-classes were immunology related.

As a different possible use of our results, note that 
some of the genes in the T-Cell related gene cluster LG5
have no determined function, and may be candidates for new T-Cell
genes. This assumption is supported both by the fact that these genes
were found to be correlated with other T-Cell genes, and by the fact
that they support the differentiation between T-ALL and B-ALL.

\subsection*{Analysis of Colon cancer data}

The data set we consider next contains  
40 colon tumor samples and 22 normal
colon samples, analyzed with an Affymetrix oligonucleotide array complementary
to more than 6500 human genes and ESTs.
Following Alon \al \cite{Alon99}, we chose to work only with the 2000 genes of
greatest minimal expression over the samples. 
We normalized the data 
to get a $2000 \times 62$ expression level matrix ${\cal A}$.

The CTWC algorithm was applied
to this data set. 97 stable gene clusters (CG1-97) and 76 stable sample clusters
(CS1-76) were obtained in two iterations. 

\begin{figure*}[tb]
  \centerline{
  \psfig{figure=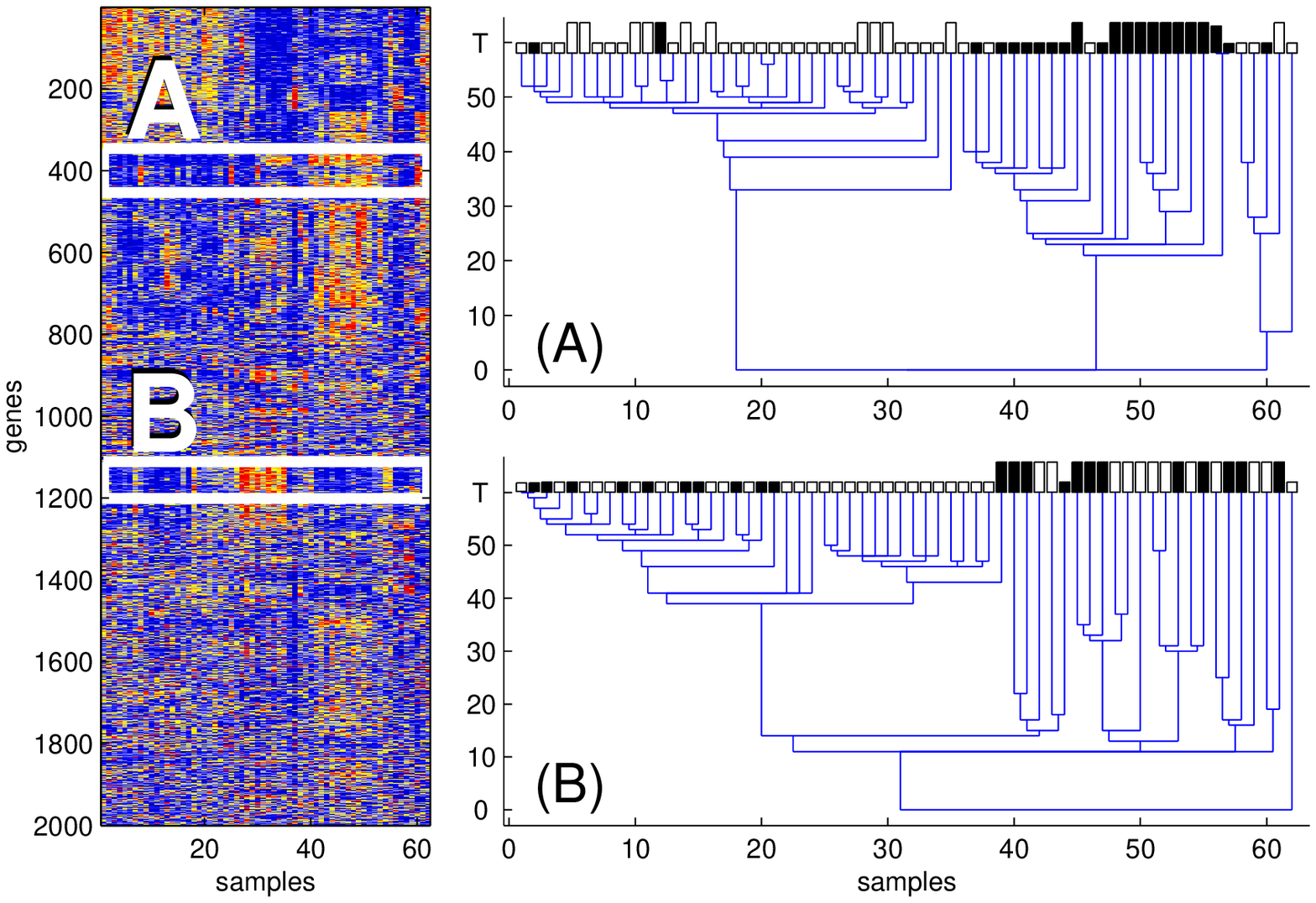,height=9.6cm}
  }
\caption{\small{The expression level matrix of the colon
experiment is shown on the left. Rows correspond to different genes,
ordered by clustering them using all the samples. 
The two boxes contain expression data of all samples for two gene clusters.
(A) Using the genes
of the first cluster, clear separation between tumor samples (white) and normal ones
(black) is obtained. (B) Another separation of the samples is obtained using the
second gene cluster. This separation is consistent with  two distinct experimental protocols, 
denoted by short and long bars.
The vertical axis is the ``temperature'' parameter $T$ and on the horizontal axis the samples
are ordered according to the dendrogram.
}}
\label{fig:colon}
\end{figure*}

{\bf Identifying genes that partition the samples 
according to a known classification. } 
Again we search first for gene clusters which differentiate the samples according 
to the known normal/tumor classification. We found
4 gene clusters (CG1-4) that partition the samples this way.
The genes of these clusters can be used if one wishes to construct a classifier
for diagnosis purposes (see Fig. \ref{fig:colon}(A)).

{\bf Discovering new partitions.} 
Five clusters of genes (CG2,CG4-CG7) generated very stable clusters of samples. 
Two of the five (CG2,CG4) differentiated 
tumor and normal; two other were less interesting since 
the clusters they generated 
contained most of the samples. The gene cluster CG5, however, 
gave rise to a clear partition of the samples into 
two clusters, of 39 and 23 tissues (see Fig. \ref{fig:colon}(B)). 
Checking with the experimentalists\footnote{U.Alon, K.Gish, D.Mack \& A.Levine,
Private communication.}
We discovered that this separation coincides almost precisely 
with a change of the experimental
protocol; 22 RNA samples were extracted using a 
poly-A detector ('protocol-A'), and the other 40 samples were prepared by
extracting total RNA from the cells ('protocol-B').
Cursory examination did not yield any obvious 
common features among the 
29 genes of the cluster CG5 that gave rise to this separation of the tissues.

\begin{figure*}[tb]
  \centerline{
  \psfig{figure=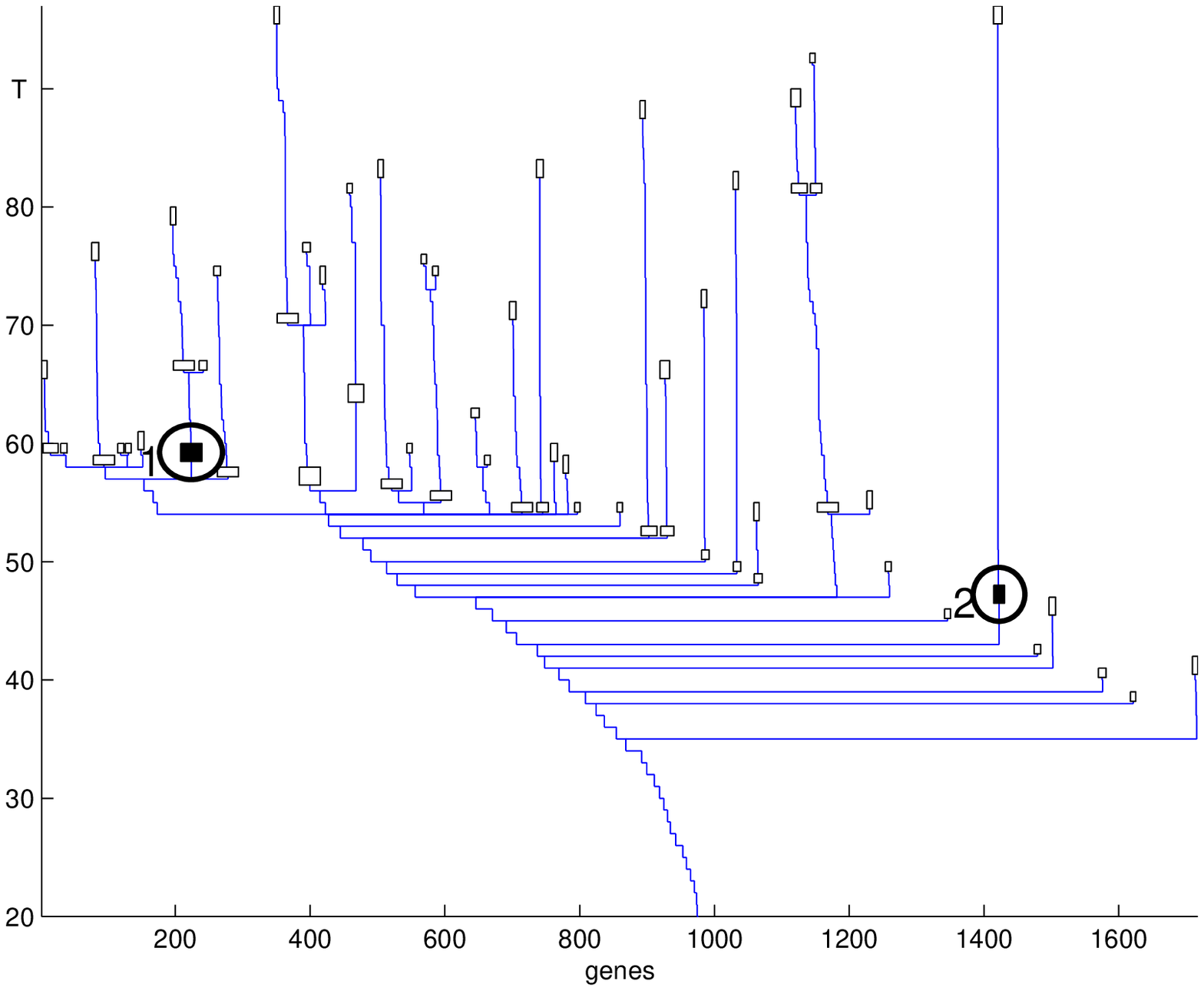,height=6.4cm}
\hspace{1cm}
  \psfig{figure=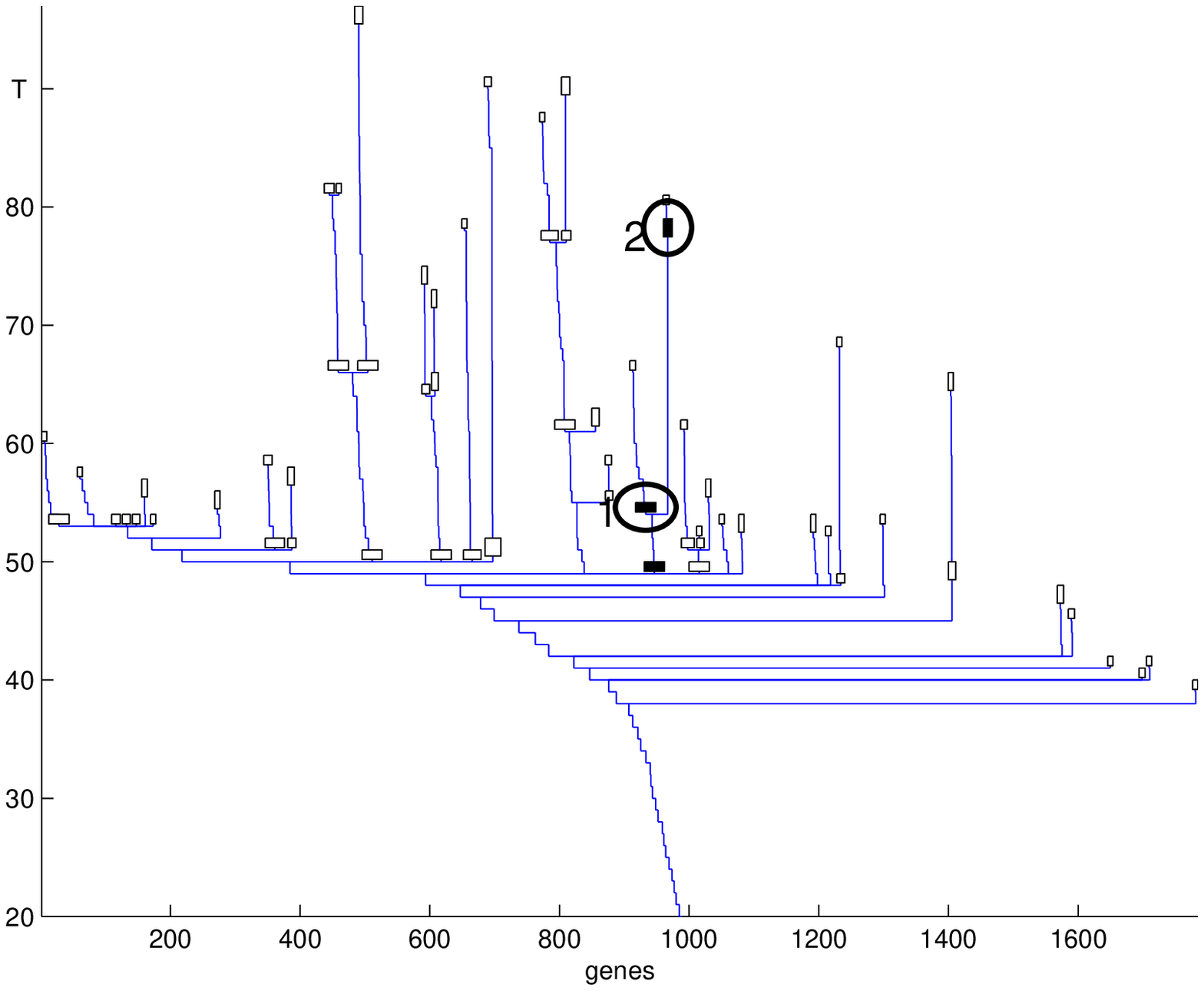,height=6.4cm}
}
\caption{\small{Clustering genes of the colon cancer experiment, (A) using all samples 
and (B) using only tumor samples as the feature sets. 
Each node of this dendrogram represents a cluster; only
clusters of size larger than 9 genes are shown. The last such clusters of
each branch, as well as non-terminal clusters that were 
selected for presentation and analysis
are shown as boxes. 
In each dendrogram the genes are ordered according to the 
corresponding cluster analysis. The two circled 
clusters of the first dendrogram
are reproduced also in the second, but there the two share a common 'parent' 
in the tree. Note that the stability of
a cluster is easily read off a dendrogram produced by the SPC algorithm.}}
\label{fig:colongenes}
\end{figure*}

{\bf Identifying conditionally correlated genes and sub-partitions}
Finally, we turn to identify conditionally correlated genes by 
comparing stable gene clusters formed when using different sample sets as 
features. We found that most gene clusters form irrespectively of the samples 
that are used. We did find, however, 4 special groups of genes (CG8-11)
that formed clear and stable clusters when using only the tumor samples as
features, but were relatively uncorrelated, i.e. spread across
the dendrogram of genes, when clustering was performed based on all 
the samples or only the normal ones.

One of these 4 clusters, (CG9), breaks up, at a higher resolution, 
into two sub-clusters, as shown in Fig. \ref{fig:colongenes}(B).
One of these sub-clusters, (CG12), consists of 51 genes, all of which are 
related to cell growth 
(ribosomal proteins and elongation factors). 
The other sub-cluster, (CG13), contains 17 genes,
many of which are related to intestinal epithelial cells 
(\eg mucin, cathespin proteases).
Interestingly, when clustering the genes on the basis of
the either all samples or only the normal ones, 
both clusters (CG12 and CG13) appear as two uncorrelated
distinct clusters, and their positions in the dendrogram  
are quite far from each other (Fig. \ref{fig:colongenes}).

The high correlation between growth genes and epithelial genes, observed in
tumor tissue, suggests that it
is the epithelial cells that are rapidly growing. In the normal samples there
is smaller correlation, indicating that the expression of growth genes  
is not especially high in the normal epithelial cells. 
These results are consistent with the epithelial origin of colon tumor.

Two other groups of genes formed clusters  only over the tumor cells.
One (CG11, of 34 genes) is related to the immune system 
(HLA genes and immunoglobulin receptors). The second
(CG10, of 62 genes) seems to be a concatenation of
genes related to epithelial cells (endothelial growth factor  
and retinoic acid), and of
muscle and nerve related genes.
We could not find any common function for the genes in 
the fourth cluster (CG8).

Clustering the genes on the basis of their expression over 
only the normal samples revealed 
three gene clusters (CG14-16) which did not form when either the entire set of
samples or the tumor tissues were used. 
Again, we could not find a clear common function for these genes. 
Each cluster contains genes that apparently take part in 
some process that takes place in normal cells, but is suppressed 
in tumor tissues.


\section*{Summary and discussion}
\label{sec:discuss}
We proposed a new method for analysis of gene microarray data. 
The main underlying idea of our method is to zero in on small subsets
of the massive expression 
patterns obtained from thousands of genes for a large number of samples.
A cellular process of interest may involve a relatively small subset of the
genes in the dataset, and 
the process may take place only in a small number of samples. Hence
when the full data set is analyzed,
the ``signal" of this process may be completely overwhelmed by 
the ``noise" generated
by the vast majority of unrelated data. 

We are looking for a relatively small group of genes, which can be used as the
features used to cluster a subset of the samples. Alternatively, we try to identify
a subset of the samples that can be used in a similar way to identify genes with
correlated expression levels.
Identifying pairs of subsets of genes and samples, which produce 
significant stable clusters in this way, is a computationally complex task.
We demonstrated that the Coupled Two-Way Clustering
technique provides an efficient method to produce such subgroups.

The CTWC algorithm provides a broad
list of stable gene and sample clusters, together with various connections among
them. This information can be used to perform the most important tasks
in microarray data analysis, such as identification of cellular processes  
and the conditions for their activation; establishing connection between 
gene groups and biological processes; and finding 
partitions of known classes of samples into sub-groups. 

We reemphasize that CTWC is applicable with any reasonable choice of
clustering algorithm, as long as it is capable of identifying stable
clusters. In this work we reported results obtained using 
the {\em super-paramagnetic} clustering algorithm (SPC),
which is especially suitable for gene microarray data analysis due
to its robustness against noise which is inherent in such experiments.

The power of the CTWC method was demonstrated on
data obtained in two gene microarray experiments.
In the first experiment the gene expression profile in
bone marrow and peripheral blood cells of
72 leukemia patients was measured using gene microarray technology.
Our main results for this data were the following:
({\it i}) The connection between T-Cell related genes and the sub-classification
of the ALL samples, into T and B-ALL, was revealed in an unsupervised fashion.
({\it ii}) We found a stable partition of the AML patients into two groups: those 
who were treated (with known results), and all others.
This partition was revealed by a cluster
of cell growth related genes. This observation may serve as a clue
for a possible use of the CTWC method in understanding the
effects of treatment.

The second experiment used gene microarray technology to probe the
gene expression profile of 40 colon tumor samples and 22 normal
colon tissues. 
Using CTWC we find a different, less obvious stable partition of the 
samples into two clusters. To find this
partition, we had to use a subset of the genes.	The new partition 
turned out to reflect two different experimental protocols.
We deduce that the genes which gave rise to
this partition of the samples are the ones which were sensitive 
to the change of protocol. 

Another result that was obtained in an unsupervised manner using CTWC, 
is the connection
between epithelial cells and the growth of cancer. When we looked 
at the expression profiles over only the tumor tissues,
a cluster of cell growth
genes was found to be highly correlated with epithelial genes. This correlation 
was absent when the normal tissues were used. 

These novel features, discovered in data sets which were previously
investigated by conventional clustering analysis, demonstrate the 
strength of CTWC. We find CTWC  to be especially useful for gene
microarray data analysis, but it may be a useful tool for investigating
other kinds of data as well.

\subsubsection*{Acknowledgments}
{\small
We thank N. Barkai for helpful discussions. The help provided
by U.Alon in all stages of this work has been invaluable; he discussed with us his
results at an early stage, provided us his data files and shared generously
his understanding and insights. This research was partially supported by
the Germany - Israel Science Foundation (GIF).
}

\end{multicols}
\end{document}